\documentstyle[12pt]{article}

\newcounter{eq}
\newcounter{sc}

\def\overleftrightarrow#1{\vbox{\ialign{##\crcr
 $\leftrightarrow$\crcr\noalign{\kern-1pt\nointerlineskip}
 $\hfil\displaystyle{#1}\hfil$\crcr}}}

\newlength{\minitwocolumn}
\begin{document}

\begin{flushright}
DPUR/TH/60\\
June, 2018\\
\end{flushright}
\vspace{20pt}

\pagestyle{empty}
\baselineskip15pt

\begin{center}
{\large\bf  Planck and Electroweak Scales Emerging from Conformal Gravity
\vskip 1mm }

\vspace{20mm}

Ichiro Oda\footnote{
           E-mail address:\ ioda@sci.u-ryukyu.ac.jp
                  }

\vspace{10mm}
           Department of Physics, Faculty of Science, University of the 
           Ryukyus,\\
           Nishihara, Okinawa 903-0213, Japan\\

\end{center}


\vspace{10mm}
\begin{abstract}

We show that both the Planck and electroweak mass scales can be generated from conformal
gravity via the Coleman-Weinberg mechanism of dimensional transmutation.  
At the first step, the Planck scale is generated via the Coleman-Weinberg mechanism in the sector 
of conformal gravity, which means that radiative corrections associated with gravitons induce 
spontaneous symmetry breakdown of a local conformal symmetry.  At the second step, the vacuum 
expectation value of a scalar field  is transmitted to the sector of the standard model via a potential
involving the conformally invariant part and the contribution from the Coleman-Weinberg mechanism, 
thereby generating the electroweak scale. The huge hierarchy between the two scales 
can be explained in terms of a very tiny coupling constant between the scalar and the Higgs field
in a consistent way.

\end{abstract}

\newpage
\pagestyle{plain}
\pagenumbering{arabic}


\rm
\section{Introduction}

The observation \cite{Atlas, CMS} of a relatively light Higgs particle with properties consistent with 
the standard model (SM) has concluded the quest to complete the particle spectrum of the SM,
but the SM itself still leaves many of unsolved questions. For instance, the crucial question remains why
the vacuum expectation value of the Higgs field and its mass are much smaller than any high scale
of new physics or the Planck scale. In particular, the non-observation of supersymmetric particles around 
the TeV scale has cast some doubts on the idea of naturalness that the Higgs boson is 
accompanied with its supersymmetric partners which protect the lightness of Higgs boson from 
the huge radiative corrections.

With the discovery of the Higgs boson and the absence of any sign of new particles at the LHC,
it seems that we are now in an era of the long overdue paradigm-shift from the view that the SM is 
an incomplete theory and should be replaced with a completely new theory like supersymmetry or
GUT to the view that the overall structure of the SM is true and we should look for its minimal
extension which largely preserves the structure of the SM. To put differently, over the past years
after the discovery of the Higgs boson at the LHC, we have learnt that the SM is equipped with
a most economical and beautiful structure and we are watching mounting evidence that its minor
extension is enough to survive to the highest energy such as the Planck energy scale where 
a theory of quantum gravity (QG) possibly unifies gravity with the other interactions in nature. 	 

In recent years, the idea that instead of supersymmetry a global scale symmetry might play an
important role in the naturalness problem has been put forward by Bardeen \cite{Bardeen} 
and afterwards various interesting models, which we call the scale invariant SM,  have been 
constructed.\footnote{There is an extensive literature on this subject, a small selection being 
\cite{Hempfling}-\cite{Englert}.} In the scenario of "great desert" mentioned above, the SM or 
physics beyond the SM (BSM) at the electroweak scale is directly unified with the gravitational theory 
at the Planck scale. Then it is natural to conjecture that the gravitational theory might also have 
a global scale symmetry as in the scale invariant SM. 

However, no-hair theorem \cite{MTW} of quantum black holes suggests that global additive conservation laws 
such as baryon and lepton number conservation cannot hold in any consistent quantum gravity theory.
Indeed, in string theory, we never get any additive conservation laws and at least in known
string vacua, the additive global symmetries turn out to be either gauge symmetries or 
explicitly violated. By contrast, gauge symmetries such as $U(1)$
electric charge conservation law cause no trouble for black hole physics. Thus, in the
gravitational theory, the global scale symmetry should be promoted to a local scale symmetry, 
so a scale invariant gravity would be replaced with conformal gravity. In this article, we would like to 
pursue such a possibility that a model involving the SM and gravity is constrained by the local conformal 
symmetry.\footnote{We have already considered such models with scale symmetries at the classical level
\cite{Oda1}-\cite{Oda3}.}  

The rest of this paper is organised as follows: In the next section, we review conformal gravity. 
In section 3, we consider the Coleman-Weinberg mechanism \cite{Coleman} in conformal gravity, and see that 
the Einstein-Hilbert action of general relativity is indeed induced via spontaneous symmetry breakdown of
the local conformal symmetry by radiative corrections associated with massive gravitons. 
In section 4,  the electroweak scale is generated by a potential affected by Coleman-Weinberg 
mechanism.  Finally, we conclude in section 5.

\section{Review of conformal gravity}

In this section, let us review some salient features of conformal gravity (or Weyl gravity). The basic building block
of conformal gravity is the conformal tensor (or Weyl tensor) $C_{\mu\nu\rho\sigma}$ which is defined as
\begin{eqnarray}
C_{\mu\nu\rho\sigma} = R_{\mu\nu\rho\sigma} - ( g_{\mu [\rho} R_{\sigma] \nu} 
- g_{\nu [\rho} R_{\sigma] \mu} ) + \frac{1}{3} g_{\mu [\rho} g_{\sigma] \nu} R,
\label{Conformal tensor}
\end{eqnarray}
where $A_{[\mu} B_{\nu]} = \frac{1}{2} ( A_\mu B_\nu - A_\nu B_\mu )$, $\mu, \nu, \cdots = 0, 1, 2, 3$,
$R_{\mu\nu\rho\sigma}, R_{\mu\nu}$ and $R$ are the Riemann tensor, the Ricci tensor and the scalar curvature,
respectively.\footnote{We will follow the conventions and notation by Misner et al. \cite{MTW}.}   
The conformal tensor shares the same symmetric properties of indices as those of the Riemann tensor 
$R_{\mu\nu\rho\sigma}$ and is in addition trace-free on all its indices. It also behaves in a very simple way
under conformal transformation\footnote{In this article, conformal transformation (or Weyl transformation) is defined 
as the ${\it local}$ scale transformation.}; $C^\mu \, _{\nu\rho\sigma} \rightarrow C^\mu \, _{\nu\rho\sigma}$ under 
$g_{\mu\nu} \rightarrow \Omega^2 (x) g_{\mu\nu}$ (notice the position of indices on $C^\mu \, _{\nu\rho\sigma}$).   

With these properties, it is easy to see that the following action of conformal gravity is invariant under the
conformal transformation
\begin{eqnarray}
S_W &=& - \frac{1}{2 \xi^2} \int d^4 x \sqrt{-g} C_{\mu\nu\rho\sigma} C^{\mu\nu\rho\sigma}
\nonumber\\
&=& - \frac{1}{2 \xi^2} \int d^4 x \sqrt{-g} ( R_{\mu\nu\rho\sigma} R^{\mu\nu\rho\sigma}
- 2 R_{\mu\nu} R^{\mu\nu} + \frac{1}{3} R^2 )
\nonumber\\
&=& - \frac{1}{\xi^2} \int d^4 x \sqrt{-g} ( R_{\mu\nu} R^{\mu\nu} - \frac{1}{3} R^2 ),
\label{Weyl action}
\end{eqnarray}
where the coupling constant $\xi$ is a dimensionless constant reflecting a conformal symmetry, and in the last
equality we have used the fact that $\sqrt{-g} ( R_{\mu\nu\rho\sigma} R^{\mu\nu\rho\sigma}
- 4 R_{\mu\nu} R^{\mu\nu} + R^2 )$ is a total derivative.
The classical field equation for conformal gravity in vacuum can be derived from the action (\ref{Weyl action}):
\begin{eqnarray}
g^{\mu\nu} \nabla^2 R - 6 \nabla^2 R^{\mu\nu} + 2 \nabla^\mu \nabla^\nu R - g^{\mu\nu} R^2
+ 3 g^{\mu\nu} R^2_{\rho\sigma} + 4 R R^{\mu\nu} - 12 R^{\mu\rho} R^\nu \, _\rho = 0,
\label{Conf-eq}
\end{eqnarray}
where $\nabla^2 \equiv g^{\mu\nu} \nabla_\mu \nabla_\nu$. We can verify that any Einstein metric, which
satisfies $R_{\mu\nu} = \Lambda g_{\mu\nu}$, is a solution to this equation. Thus all the classical solutions 
of Einstein gravity in vacuum are also solutions of conformal gravity.

In the weak field approximation we can expand  
\begin{eqnarray}
g_{\mu\nu} = \eta_{\mu\nu} + \xi h_{\mu\nu},
\label{Metric-fluct}
\end{eqnarray}
where $\eta_{\mu\nu}$ is a flat Minkowski metric and $| h_{\mu\nu} | \ll 1$. Then, up to quadratic terms in
$h_{\mu\nu}$ the action (\ref{Weyl action}) can be cast to
\begin{eqnarray}
S_W = - \frac{1}{4} \int d^4 x h^{\mu\nu} P^{(2)}_{\mu\nu, \rho\sigma} \Box^2 h^{\rho\sigma},
\label{Weyl action2}
\end{eqnarray}
where $P^{(2)}_{\mu\nu, \rho\sigma}$ is the projection operator for spin-2 modes\footnote{We have used the
definitions of projection operators in \cite{Oda4, Oda5}.} and $\Box \equiv 
\eta^{\mu\nu} \partial_\mu \partial_\nu$. The linearized equations of motion therefore become
\begin{eqnarray}
P^{(2)}_{\mu\nu, \rho\sigma} \Box^2 h^{\rho\sigma} = 0.
\label{Linear-eq}
\end{eqnarray}
It is known that the linearized equations of motion for conformal gravity possess six physical degrees of
freedom which consist of massless spin-2 and spin-1 normal modes and a massless spin-2 ghost mode
\cite{Lee, Riegert}.

The advantages of conformal gravity compared to Einstein gravity are that conformal gravity is renormalizable
\cite{Stelle} and asymptotically free \cite{Tonin, Fradkin} so that it is a ultra-violet (UV) complete 
quantum field theory. Moreover, the square of conformal tensor, which appears in the action of conformal gravity,
is certainly positive definite in the Euclidean signature. For each topological class, the action is therefore bounded from
below and the saddle-point approximation can be applied in order to understand non-perturbative phenomena
such as instantons \cite{Strominger}. 

However, the price we have to pay is the presence of a ghost violating the unitarity.
In this article, we will not address the issue of the ghost in detail but briefly comment on it in section 4.\footnote{For recent 
progress on this topics, see a review article \cite{Salvio}. This article also includes a full set of the one-loop
renormalization group equations for the conformal SM.}   Another problem of conformal gravity is related to conformal 
anomaly \cite{Duff1, Duff2}. It has been thought for a long time that conformal gravity is not a consistent quantum
field theory owing to the existence of conformal anomaly. However, it is nowadays an established fact that there exists 
a quantization procedure of preserving the local conformal symmetry at the quantum level though the theory 
in essence becomes non-renormalizable \cite{Englert2}-\cite{Ghilencea3}.\footnote{The renormalizability 
is not essential for the validity on this point.}   In this sense, we do not have to 
pay special attention to the issue of conformal anomaly any longer. As a last problem, 
on the phenomenological grounds, the macroscopic behavior of the gravitational field is governed by 
the Einstein-Hilbert action, and not the one of conformal gravity. Thus, if quantum conformal gravity 
has any physical relevance, its effective action must include the induced Einstein-Hilbert action. 
It is natural to conjecture that the Einstein-Hilbert term arises when radiative corrections break the local conformal 
symmetry. In the next section, we will explicitly show that this is indeed the case.

To close this section, let us comment on the conformally invariant scalar-tensor gravity whose action reads
\cite{Dirac}
\begin{eqnarray}
S_{ST} = \int d^4 x \sqrt{-g} \left( \frac{1}{12} \phi^2 R + \frac{1}{2} g^{\mu\nu} \partial_\mu \phi 
\partial_\nu \phi \right),
\label{Conf-ST-action}
\end{eqnarray}
where $\phi$ is a real scalar field.\footnote{The generalization to a complex 
scalar field is straightforward as seen later.}  This action is invariant under conformal transformation
\begin{eqnarray}
g_{\mu\nu} \rightarrow \Omega^2 (x) g_{\mu\nu},  \quad 
\phi \rightarrow \Omega^{-1} (x) \phi.
\label{Conf transf}
\end{eqnarray}
Note that the scalar field $\phi$ is not a normal field but a ghost. However, this ghost can be eliminated
by taking the gauge condition for conformal transformation\footnote{Alternatively, this gauge fixing
procedure can be interpreted as a field redefinition from the conformally invariant metric
$g^\prime_{\mu\nu} = \frac{1}{6 M^2_{pl}} \phi^2 g_{\mu\nu}$ to the metric $g_{\mu\nu}$.}
\begin{eqnarray}
\phi (x) = \sqrt{6} M_{pl},
\label{Gauge-Conf transf}
\end{eqnarray}
with $M_{pl}$ being the reduced Planck mass, and this gauge condition reduces the action (\ref{Conf-ST-action})
to the conventional Einstein-Hilbert action
\begin{eqnarray}
S_{EH} = \frac{M^2_{pl}}{2}  \int d^4 x \sqrt{-g} R.
\label{EH-action}
\end{eqnarray}
Accordingly, when a real scalar field $\phi$ coexists with gravity, the most general action which is conformally invariant, is
of form
\begin{eqnarray}
S_C = \int d^4 x \sqrt{-g} \left( - \frac{1}{2 \xi^2} C_{\mu\nu\rho\sigma} C^{\mu\nu\rho\sigma}
+ \frac{1}{12} \phi^2 R + \frac{1}{2} g^{\mu\nu} \partial_\mu \phi \partial_\nu \phi \right).
\label{Conf-action}
\end{eqnarray}

\section{Coleman-Weinberg mechanism in conformal gravity}

The standard model (SM) of particle physics without the Higgs mass is known to have an extra symmetry,
which is a $\it{global}$ scale symmetry. This global scale symmetry can be enlarged to a $\it{local}$ scale symmetry,
which we call $\it{conformal \, symmetry}$ in this article, by introducing the conformally invariant coupling
with gravity. In this section, we therefore wish to work with this conformally invariant SM with the more general,
conformally invariant gravity action (\ref{Conf-action}), and see how the Coleman-Weinberg mechanism 
generates the Einstein-Hilbert action through spontaneous symmetry breaking of conformal symmetry by
radiative corrections associated with gravitational fluctuations. 

The Lagrangian density we study is 
\begin{eqnarray}
\frac{1}{\sqrt{-g}} {\cal{L}}_C &=& - \frac{1}{2 \xi^2} C_{\mu\nu\rho\sigma} C^{\mu\nu\rho\sigma}
+ \frac{1}{12} \phi^2 R + \frac{1}{2} g^{\mu\nu} \partial_\mu \phi \partial_\nu \phi
- \frac{1}{6} (H^\dagger H) R - g^{\mu\nu} (D_\mu H)^\dagger (D_\nu H)
\nonumber\\
&+& V(\phi, H) + L_m,
\label{Conf-SM-action}
\end{eqnarray}
where $H$ is the Higgs doublet, $D_\mu$ is a covariant derivative including the SM gauge fields, and
$L_m$ denotes the remaining Lagrangian density of the SM without the Higgs mass term. Moreover,
the new potential $V(\phi, H)$ beyond the SM, which is conformally invariant, is added and has the form
\begin{eqnarray}
V(\phi, H) = \frac{\lambda_\phi}{4 !} \phi^4  + \lambda_{H \phi} (H^\dagger H) \phi^2  
+ \frac{\lambda_H}{2} (H^\dagger H)^2,
\label{V(phi, H)}
\end{eqnarray}
where all the coupling constants $\lambda_\phi, \lambda_{H \phi}$ and $\lambda_H$ are dimensionless
and we will assume that $\lambda_{H \phi} < 0$ and $| \lambda_{H \phi} | \ll 1$. The physical 
meaning of these two assumptions will be mentioned in the next section.
Since we assume $| \lambda_{H \phi} | \ll 1$, we can envision the process of symmetry breaking
as two independent steps. At the first step of symmetry breaking, the Coleman-Weinberg mechanism 
would generate a vacuum expectation value (VEV) for the scalar field $\phi$ around the Planck scale.
Then, at the second step, it is expected that this huge VEV is transmitted to the SM sector through
the full form of the potential $V(\phi, H)$, thereby generating the Higgs mass term. In this section, 
we will focus on the first step of symmetry breaking.  

To this aim, let us first expand the scalar field and the metric around a classical field $\phi_c$ and 
a flat Minkowski metric $\eta_{\mu\nu}$ like
\begin{eqnarray}
\phi = \phi_c + \varphi,    \quad 
g_{\mu\nu} = \eta_{\mu\nu} + \xi h_{\mu\nu},
\label{Expansion}
\end{eqnarray}
where we take $\phi_c$ to be a constant since we are interested in the effective potential depending on
the constant $\phi_c$. Then, as seen in Eq. (\ref{Weyl action2}), the Lagrangian density corresponding to
the action $S_W$ takes the form
\begin{eqnarray}
{\cal L}_W = - \frac{1}{4} h^{\mu\nu} P^{(2)}_{\mu\nu, \rho\sigma} \Box^2 h^{\rho\sigma},
\label{Weyl Lagr}
\end{eqnarray}
and in a similar manner,  up to quadratic terms in $h_{\mu\nu}$, the Lagrangian density corresponding to 
the action $S_{ST}$ in Eq. (\ref{Conf-ST-action}) reads
\begin{eqnarray}
{\cal L}_{ST} = - \frac{1}{6} \xi \phi_c \varphi \left( \eta_{\mu\nu} 
- \frac{1}{\Box} \partial_\mu \partial_\nu \right) \Box h^{\mu\nu}
+ \frac{1}{48} \xi^2 \phi^2_c h^{\mu\nu} \left( P^{(2)}_{\mu\nu, \rho\sigma} 
- 2 P^{(0, s)}_{\mu\nu, \rho\sigma} \right) \Box h^{\rho\sigma} - \frac{1}{2} \varphi \Box \varphi.
\label{Conf-ST-Lagr}
\end{eqnarray}
Hence, adding the two Lagrangian densities, we have
\begin{eqnarray}
{\cal L}_W + {\cal L}_{ST} &=& \frac{1}{4} h^{\mu\nu} \left[ \left( - \Box + \frac{1}{12} \xi^2 \phi^2_c \right)
P^{(2)}_{\mu\nu, \rho\sigma} - \frac{1}{6} \xi^2 \phi^2_c P^{(0, s)}_{\mu\nu, \rho\sigma} \right]
\Box h^{\rho\sigma}
\nonumber\\
&-& \frac{1}{6} \xi \phi_c \varphi \left( \eta_{\mu\nu} 
- \frac{1}{\Box} \partial_\mu \partial_\nu \right) \Box h^{\mu\nu} - \frac{1}{2} \varphi \Box \varphi.
\label{Total-Lagr}
\end{eqnarray}

Here let us set up the gauge-fixing conditions. For diffeomorphisms, we adopt the gauge condition
\begin{eqnarray}
\chi_\mu \equiv \partial^\nu ( h_{\mu\nu} - \frac{1}{4} \eta_{\mu\nu} h ) = 0,
\label{Gauge-diffo}
\end{eqnarray}
which is invariant under conformal transformation. The Lagrangian density for this gauge condition
and its corresponding FP ghost term is given by 
\begin{eqnarray}
{\cal L}_{GF + FP} = - \frac{1}{2 \alpha} \chi_\mu \chi^\mu + \bar c_\mu ( \Box \eta^{\mu\nu}
+ \frac{1}{2} \partial^\mu \partial^\nu ) c_\nu,
\label{Gauge+FP}
\end{eqnarray}
where $\alpha$ is a gauge parameter. Let us recall that in case of the higher derivative gravity, 
one usually works with a more general gauge-fixing term ${\cal L}_{GF} = \chi_\mu Y^{\mu\nu} 
\chi_\nu$ where $Y^{\mu\nu}$ is the weight function involving derivatives and gauge parameters
\cite{Fradkin}.  However, for the present purpose it turns out to be sufficient to choose 
$Y^{\mu\nu} = - \frac{1}{2 \alpha} \eta^{\mu\nu}$. 
Next, we fix the gauge freedom corresponding to conformal transformation by the gauge
condition
\begin{eqnarray}
h \equiv \eta^{\mu\nu} h_{\mu\nu}  = 0.
\label{Gauge-conf}
\end{eqnarray}
Since this gauge fixing condition and the conformal transformation contain no derivatives, one can 
neglect the corresponding FP ghost term in the one-loop approximation. 

Consequently, we can obtain a quantum Lagrangian density up to quadratic terms in $h_{\mu\nu}$:
\begin{eqnarray}
{\cal L}_W + {\cal L}_{ST} + {\cal L}_{GF + FP} &=& \frac{1}{4} h^{\mu\nu} \left( - \Box 
+ \frac{1}{12} \xi^2 \phi^2_c \right) P^{(2)}_{\mu\nu, \rho\sigma} 
\Box h^{\rho\sigma} - \frac{1}{2} \varphi^\prime \Box \varphi^\prime
\nonumber\\
&-& \frac{1}{2 \alpha} ( \partial^\nu h_{\mu\nu} )^2 -  \bar c_\mu ( \Box \eta^{\mu\nu}
+ \frac{1}{2} \partial^\mu \partial^\nu ) c_\nu,
\label{Q-Lagr}
\end{eqnarray}
where we have simplified this expression by using the gauge condition (\ref{Gauge-conf}) and introducing 
$\varphi^\prime$ defined as
\begin{eqnarray}
\varphi^\prime \equiv \varphi - \frac{1}{6} \xi \phi_c \Box^{-1} \partial_\mu \partial_\nu h^{\mu\nu},
\label{varphi-prime}
\end{eqnarray}
which corresponds to the massless Nambu-Goldstone boson.
Based on this quantum Lagrangian density (\ref{Q-Lagr}), we can evaluate the one-loop effective action
by integrating out quantum fluctuations associated with $h_{\mu\nu}$.  Then, up to a classical potential,
the effective action $\Gamma [\phi_c]$ reads
\begin{eqnarray}
\Gamma [\phi_c] = i \frac{5}{2} \log \mathrm{det} \left( - \Box + \frac{1}{12} \xi^2 \phi^2_c \right).
\label{EA}
\end{eqnarray}
In this expression, the factor $5$ comes from the fact that a massive spin-2 state possesses 
five physical degrees of freedom and we have ignored the part of the effective action which is
independent of $\phi_c$.

To calculate $\Gamma [\phi_c]$, we can proceed by following the same line of the arguments as in \cite{Peskin}. 
First of all, let us note that $\Gamma [\phi_c]$ can be rewritten as follows:
\begin{eqnarray}
\Gamma [\phi_c] &=& i \frac{5}{2} \mathrm{Tr} \log \left( - \Box + \frac{1}{12} \xi^2 \phi^2_c \right)
\nonumber\\
&=& i \frac{5}{2} \sum_k \log \left( k^2 + \frac{1}{12} \xi^2 \phi^2_c \right)
\nonumber\\
&=& (VT) i \frac{5}{2} \int \frac{d^4 k}{(2 \pi)^4} \log \left( k^2 + \frac{1}{12} \xi^2 \phi^2_c \right)
\nonumber\\
&=& (VT) \frac{5}{2} \frac{\Gamma(- \frac{d}{2})}{(4 \pi)^{\frac{d}{2}}} 
\left( \frac{1}{12} \xi^2 \phi^2_c \right)^{\frac{d}{2}},
\label{EA2}
\end{eqnarray}
where $(VT)$ denotes the space-time volume and in the last equality we have used the Wick rotation and 
the dimensional regularization.

Next, recall that the conventional method to calculate the effective potential is to introduce the counter-terms to
subtract UV-divergences and then impose the renormalization conditions to fix the finite part.
However, if we want to visualize the modification of the lowest-order results which is generated by
radiative corrections, we can apply some renormalization scheme which can be implemented more
easily \cite{Peskin}. For instance, the minimal subtraction scheme is simply to remove the $\frac{1}{\epsilon}$
poles (where $\epsilon \equiv 4 - d$) in divergent expressions. In this article, we will adopt the
intermediate method; we first subtract the $\frac{1}{\epsilon}$ poles and then fix the finite part by
imposing the renormalization conditions.  By subtracting the $\frac{1}{\epsilon}$ poles, the effective
potential $V_{eff} (\phi_c)$ in the one-loop approximation becomes
\begin{eqnarray}
V_{eff} (\phi_c) &\equiv& - \frac{1}{VT} \Gamma [\phi_c] 
\nonumber\\
&=& \frac{5}{9216 \pi^2} \xi^4 \phi^4_c \left( \log \frac{\phi^2_c}{\mu^2} + c \right),
\label{EP}
\end{eqnarray}
where $\mu$ is the renormalization mass scale and $c$ is a constant to be determined by the renormalization
conditions:
\begin{eqnarray}
\left. V_{eff} \right\vert_{\phi_c = 0} = \left. \frac{d^2 V_{eff}}{d \phi^2_c} \right\vert_{\phi_c = 0} 
= \left. \frac{d^4 V_{eff}}{d \phi^4_c} \right\vert_{\phi_c = \mu} = 0.
\label{Ren-cond}
\end{eqnarray}
As a result, we have the effective potential
\begin{eqnarray}
V_{eff} (\phi_c) = \frac{5}{9216 \pi^2} \xi^4 \phi^4_c \left( \log \frac{\phi^2_c}{\mu^2} - \frac{25}{6} \right).
\label{EP2}
\end{eqnarray}
Finally, by adding the classical potential we can arrive at the effective potential in the one-loop approximation
\begin{eqnarray}
V_{eff} (\phi_c) = \frac{\lambda_\phi}{4 !} \phi^4_c + \frac{5}{9216 \pi^2} \xi^4 \phi^4_c 
\left( \log \frac{\phi^2_c}{\mu^2} - \frac{25}{6} \right).
\label{EP3}
\end{eqnarray}

It is easy to see that this effective potential has a minimum at $\phi_c = \langle \phi \rangle$ away from the origin 
where the effective potential, $V_{eff} (\langle \phi \rangle)$, is negative. Since the renormalization mass $\mu$ 
is arbitrary, we will choose it to be the actual location of the minimum, $\mu = \langle \phi \rangle$ \cite{Coleman}:
\begin{eqnarray}
V_{eff} (\phi_c) = \frac{\lambda_\phi}{4 !} \phi^4_c + \frac{5}{9216 \pi^2} \xi^4 \phi^4_c 
\left( \log \frac{\phi^2_c}{\langle \phi \rangle^2} - \frac{25}{6} \right).
\label{EP4}
\end{eqnarray}
Since $\phi_c = \langle \phi \rangle$ is defined to be the minimum of $V_{eff}$, we deduce
\begin{eqnarray}
0 &=& \left. \frac{d V_{eff}}{d \phi_c} \right\vert_{\phi_c = \langle \phi \rangle}
\nonumber\\ 
&=& \left( \frac{\lambda_\phi}{6} - \frac{55}{6912 \pi^2} \xi^4 \right) \langle \phi \rangle^3,
\label{Min-cond}
\end{eqnarray}
or equivalently, 
\begin{eqnarray}
\lambda_\phi = \frac{55}{1152 \pi^2} \xi^4.
\label{Min-cond2}
\end{eqnarray}
The substitution of Eq. (\ref{Min-cond2}) into $V_{eff}$ in (\ref{EP4}) leads to
\begin{eqnarray}
V_{eff} (\phi_c) = \frac{5}{9216 \pi^2} \xi^4 \phi^4_c 
\left( \log \frac{\phi^2_c}{\langle \phi \rangle^2} - \frac{1}{2} \right).
\label{EP5}
\end{eqnarray}
Thus, the effective potential is now parametrized in terms of $\xi$ and $\langle \phi \rangle$
instead of $\xi$ and $\lambda_\phi$; it is nothing but dimensional transmutation, i.e., a dimensionless
coupling constant $\lambda_\phi$ is traded for a dimensional quantity $\langle \phi \rangle$ via
spontaneous symmetry breakdown of the $\it{local}$ conformal symmetry.   

With $\phi = \langle \phi \rangle + \varphi$, the Lagrangian density of the conformally invariant
scalar-tensor gravity plus the classical potential term produces the Einstein-Hilbert term and 
a massive scalar ghost like
\begin{eqnarray}
\frac{1}{\sqrt{-g}} {\cal L}^\prime &\equiv& \frac{1}{12} \phi^2 R 
+ \frac{1}{2} g^{\mu\nu} \partial_\mu \phi \partial_\nu \phi  
- \frac{\lambda_\phi}{4 !} \phi^4
\nonumber\\
&=& \frac{1}{12} \langle \phi \rangle^2 R 
+ \frac{1}{2} g^{\mu\nu} \partial_\mu \varphi \partial_\nu \varphi  
- \frac{\lambda_\phi}{4} \langle \phi \rangle^2 \varphi^2 + \cdots
\nonumber\\
&=& \frac{M^2_{pl}}{2} R + \frac{1}{2} g^{\mu\nu} \partial_\mu \varphi \partial_\nu \varphi  
- \frac{1}{2} m^2_\varphi \varphi^2 + \cdots,
\label{L-prime}
\end{eqnarray}
where the elipses stand for the higher-order interaction terms, and the reduced Planck mass and 
the mass of a scalar ghost are respectively defined as 
\begin{eqnarray}
M^2_{pl} =  \frac{1}{6} \langle \phi \rangle^2,   \quad m^2_\varphi = \frac{\lambda_\phi}{2} 
\langle \phi \rangle^2.
\label{Mass}
\end{eqnarray}
Eq. (\ref{Mass}) clearly shows that the minimum $\phi = \langle \phi \rangle$ of the effective potential 
is located near the Planck scale, and if we suppose $\lambda_\phi \sim {\cal O}(0.1)$ the mass of 
the scalar ghost is of order of the Planck mass scale. In other words, in our model
spontaneous symmetry breaking of the conformal symmetry happens around the Planck scale
via the Coleman-Weinberg mechanism in a natural manner.

As a final remark, in the above we have mentioned that conformal symmetry was spontaneously broken via
the Coleman-Weinberg mechanism, but precisely speaking this is not spontaneous but explicit symmetry
breaking of the conformal symmetry. This can been certified in the fact that the scalar $\varphi$ becomes 
massive because of radiative corrections as in Eq.  (\ref{L-prime}) although it is massless at the tree level 
as in Eq.  (\ref{Q-Lagr}). (In this respect, the difference between $\varphi^\prime$ and $\varphi$ 
makes no sense.) In order to have the true spontaneous symmetry breaking of the conformal symmetry, 
we should adopt a manifestly conformal invariant regularization where the renormalization scale $\mu$ 
is promoted to a dynamical field $\mu = z \phi$ with $z$ being a constant \cite{Englert2}-\cite{Ghilencea3}.

\section{Emergence of the electroweak scale}

Including the Higgs sector, after spontaneous symmetry breakdown of conformal symmetry, 
we have now an effective potential
\begin{eqnarray}
V_{eff} (\phi, H) = \frac{5}{9216 \pi^2} \xi^4 \phi^4 \left( \log \frac{\phi^2}{\langle \phi \rangle^2} 
- \frac{1}{2} \right) + \lambda_{H \phi} (H^\dagger H) \phi^2  
+ \frac{\lambda_H}{2} (H^\dagger H)^2.
\label{H-EP}
\end{eqnarray}
Inserting the minimum $\phi = \langle \phi \rangle$ to Eq. (\ref{H-EP}) and completing the square, 
the effective potential reduces to
\begin{eqnarray}
V_{eff} (\langle \phi \rangle, H) = \frac{\lambda_H}{2} \left( H^\dagger H 
+ \frac{\lambda_{H \phi}}{\lambda_H} \langle \phi \rangle^2 \right)^2
- \frac{1}{2} \left( \frac{\lambda^2_{H \phi}}{\lambda_H} + \frac{5}{9216 \pi^2} \xi^4 \right)
\langle \phi \rangle^4.
\label{H-EP2}
\end{eqnarray}
Owing to $\lambda_H > 0$, this potential has a minimum at $H^\dagger H = -\frac{\lambda_{H \phi}}{\lambda_H} 
\langle \phi \rangle^2$. Taking the unitary gauge $H^T = \frac{1}{\sqrt{2}} (0, v + h)$, this fact implies that
the square of the VEV $v$ and the Higgs mass $m_h$ is given by
\begin{eqnarray}
v^2 = \frac{2 | \lambda_{H \phi} |}{\lambda_H} \langle \phi \rangle^2,  \quad 
m_h^2 = \lambda_H v^2.
\label{v and H-mass}
\end{eqnarray}
Using Eqs.  (\ref{Mass}) and  (\ref{v and H-mass}), the magnitude of the coupling constant $\lambda_{H \phi}$
reads
\begin{eqnarray}
| \lambda_{H \phi} |  = \frac{1}{12} \left( \frac{m_h}{M_{pl}} \right)^2 \sim {\cal O} (10^{-33}).
\label{lambda}
\end{eqnarray}
This relation certainly supports our previous assumption $| \lambda_{H \phi} | \ll 1$.

At this stage, it is worthwhile to recall and reflect what we have done so far.  One appealing point in our
formalism is that starting with a conformally invariant gravity, we have generated the Einstein-Hilbert action 
via the Coleman-Weinberg mechanism at the one-loop level. In the starting Lagrangian density (\ref{Conf-SM-action}), 
the scalar field $\phi$ is a ghost with opposite sign kinetic term. The difficulty associated with (\ref{Conf-SM-action}) 
is then that, with both normal and ghost fields, the energy of the theory is unbounded from below, and therefore 
the vacuum $\langle 0| \phi | 0 \rangle = 0$ is unstable. Vacuum decay will inevitably occur at the quantum
level.  However, radiative corrections coming from massive ghost-like gravitons with spin $2$ tame this
instability and as a result the Einstein-Hilbert term describing massless normal gravitons with spin $2$
is generated.  

However, in the process of spontaneous symmetry breakdown of conformal symmetry, 
we still get a massive scalar ghost $\varphi$. Since this ghost field is in essence a conformal mode of the
graviton, by following the pragmatic attitude in \cite{Gibbons, Mazur}, let us perform the Wick rotation
over the conformal mode, $\varphi \rightarrow i \varphi$, or equivalently, $\phi \rightarrow i \phi$, thereby the ghost 
becoming a tachyon. (Here note that the minimum $\langle \phi \rangle$ is fixed in such a way that 
the square of the reduced Planck mass is positive as seen in Eq. (\ref{Mass}).) 
At the same time, the classical potential is changed as
\begin{eqnarray}
V(\phi, H) = \frac{\lambda_\phi}{4 !} \phi^4  + \lambda_{H \phi} (H^\dagger H) \phi^2  
+ \frac{\lambda_H}{2} (H^\dagger H)^2
\rightarrow \frac{\lambda_\phi}{4 !} \phi^4  - \lambda_{H \phi} (H^\dagger H) \phi^2  
+ \frac{\lambda_H}{2} (H^\dagger H)^2,
\label{V(phi, H) 2}
\end{eqnarray}
where only the sign in front of the second term is modified. Then, the positivity of this potential 
requires us to select $\lambda_{H \phi} < 0$. Thus, in this interpretation on the basis of the Wick
rotation, we can explain in a natural way why we had to choose $\lambda_{H \phi} < 0$ at the beginning.
Of course, it is not clear at present that we could perform the Wick rotation of the conformal factor of the
graviton.  Nevertheless, at least the present consideration clarifies the reason of why $\lambda_{H \phi} < 0$
is.  Moreover, it is suggested that the scalar field $\phi$, which appears in the classical potential, 
is not a normal field but might be a ghost or tachyon.

\section{Discussion}

In this article, we have discussed a locally scale invariant standard model which includes conformal
gravity and the conformally invariant scalar-tensor gravity in addition to the SM but the Higgs mass term. 
One of the interesting aspects in the model at hand is that the whole system is constrained 
by gauge symmetries and in fact it is invariant under not a global but a local conformal symmetry. 
The enlargement of symmetry from the global scale symmetry to the local conformal one is expected 
from the consideration of the no-hair theorem of quantum black holes.

The main purpose behind the present study is to understand the hierarchy problems, in particular,
the gauge hierarchy problem and the cosmological constant problem based on conformal
symmetry instead of supersymmetry.  However, to do so, the conformal symmetry must be maintained
at the quantum level in all orders of perturbation theory.  Since our world is not conformally invariant at
least in the low energy regime, the conformal symmetry must be broken spontaneously, thereby leading
to a massless dilaton. Note that in this case the advantage of the conformal invariance is preserved 
at the quantum level and the quantum conformal symmetry could play an important role in solving the
hierarchy problems. As a future work, we wish to construct such a model where the conformal
symmetry is spontaneously broken by using the manifestly conformal invariant regularization. 

Another appealing point of our formalism is that starting with conformal gravity Einstein's general
relativity is induced through radiative corrections associated with massive ghost-like gravitons.
Even if we have performed this derivation in a flat Minkowski background, it would be possible to
perform a similar calculation in a general curved background.  
 
Of course, there is a big problem to be solved in future; the issue of ghosts. If the Wick rotation for
the conformal factor were allowed, the scalar ghost would be transformed to a benign tachyon, which triggers
spontaneous symmetry breakdown at the electroweak scale. But there still remain ghosts, i.e., massive
ghost-like gravitons with spin $2$. In the one-loop level, the ghosts supply us the non-trivial VEV for the
scalar field $\phi$ through quantum effects. It is of interest to investigate whether this situation 
holds in the higher-order levels as well.

\begin{flushleft}
{\bf Acknowledgements}
\end{flushleft}
The work was supported by JSPS KAKENHI Grant Number 16K05327.


\end{document}